\documentclass[fleqn,twoside]{article}
\usepackage{espcrc2}

\usepackage{graphicx}
\usepackage[figuresright]{rotating}

\newcommand{\AmS}{{\protect\the\textfont2
  A\kern-.1667em\lower.5ex\hbox{M}\kern-.125emS}}

\title{Status, performance, and first results of the IceTop array}
\author{Todor Stanev for the IceCube Collaboration
\address{Bartol Research Institute, Department of Physics
 and Astronomy, University of Delaware, Newark, DE 19716, U.S.A.}
        \thanks{The author acknowledges the support of the
         National Science Foundation of US NSF ANT-0602679.
}}
\begin{document}
\begin{abstract}
 We describe the design and performance of IceTop, the air shower array
 on top of the IceCube neutrino detector. After the 2008/09 
 antarctic summer season both detectors are deployed at almost 3/4
 of their design size. With the current IceTop 59 stations we can
 start the study of showers of energy well above 10$^{17}$ eV.
 The paper also describes the first results from IceTop and our plans 
 to study the cosmic ray composition using several different 
 types of analysis.
\vspace{1pc}
\end{abstract}
%
\maketitle

\section{Introduction}
 IceTop is the air shower array part of the IceCube neutrino
 telescope~\cite{Tom07}.
 It supports IceCube in terms of relative and
 absolute pointing and guard against high energy air showers being
 misreconstructed as ultrahigh high energy neutrinos.
 With nearly 1 km$^2$ area and atmospheric depth of 690 g/cm$^2$
 IceTop is also an excellent tool for cosmic ray physics research. 

 The combination of IceTop with the InIce
 muon detector operates as huge three dimensional muon telescope
 that has  big advantages in studies of the cosmic ray spectrum
 and composition. This device can
 \begin{itemize} 
 \item measure the cosmic ray energy spectrum by studies of the
shower hits in the surface detectors
 \item measure the cosmic ray spectrum and composition by
studies of the angular distribution of the muon bundles in the
in-ice array.
 \item measure the spectrum and composition using coincident
 IceTop/InIce events
\end{itemize}

 IceCube can not measure the number of TeV muons
 under ice. It, however, measures the energy released
 by these muons in their propagation in 1 km of deep ice.
 The average surface energy of vertical muons that reach
 IceCube with sufficient energy to create signals 
 is slightly higher than 500 GeV.

 We briefly discuss the design and current status of IceTop,
 its calibration and performance, early air shower results,
 and outline the possible cosmic ray measurements with this
 detector.

\section{IceTop design}

 IceTop will consist of 80 stations that are located at an
 average distance of 125 meters from each other close to 
 the top of each IceCube string. Each IceTop
 station consists of two frozen water tanks. IceTop tanks are
 plastic tanks with white diffusive reflecting inner surface of
 radius 0.93 m and height of 1 m. The freezing is a long complicated
 process because the water in the tank has to be freed of the
 dissolved air which after freezing creates bubbles that make 
 the tank volume not uniform in optical clarity.
 In spite of the average antarctic 
 summer temperature of -25$^o$C,  the freezing usually takes 
 about two months.

 Each tank is equipped with two digital optical modules (DOMs)
 identical to those used by IceCube in-ice array that collect the
 Cherenkov light emitted by the shower particles that enter
 the tank. Note that the shower gamma rays convert in the tank
 and the experiment measures the energy flow of the
 extensive air shower (EAS).

 A DOMs contain a 10''  Hamamatsu R7081-2 photomultiplier tube, onboard 
 high voltage converter, flashers and onboard digitization
 electronics: two 420 ns, 300 MHz chips, each with four channels
 with different gains~\cite{IceCube_NIM}.
 The two tank PMTs are run on different gains,
 one on high (5$\times$10$^6$) gain (HG) and the other on low (10$^5$)
 gain (LG) for better dynamic range.

 Events with signal above threshold only in one tank of the
 station are considered hits by individual cosmic ray electrons,
 gamma rays, muons  or hadrons. Coincident hits in both tanks
 of a station are considered air shower events.

\section{Status of IceTop in 2008}

 The deployment of IceTop tanks goes simultaneously with that
 of InIce strings. 
 In the beginning of 2008 both IceTop and IceCube had 40 stations
 and strings. Nineteen new stations and strings were deployed during 
 the most recent season and this year the experiment will be almost
 3/4 complete. 

\begin{figure}[htb]
 \vspace*{-15pt}
\begin{center}
\includegraphics[width=78truemm]{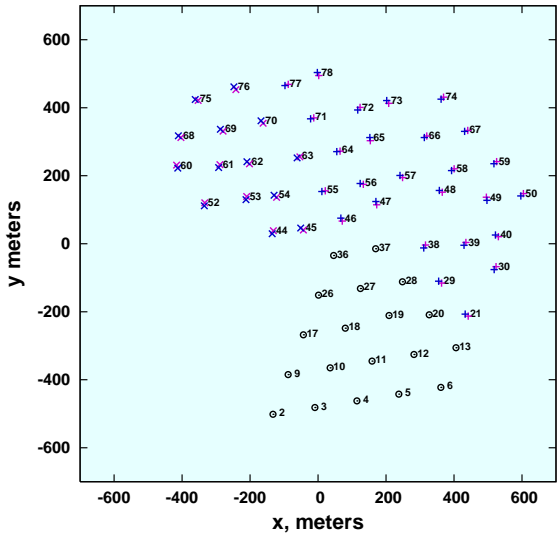}
 \vspace*{-20pt}
\caption{Map of the IceTop array. One sees tanks in stations deployed
 before the 2007/08 season (crosses), during that season (x's) 
 and in the 2008/09 season, where only the planned positions 
 of the stations are shown with circles.
\label{it40}}
\end{center}
 \vspace{-25pt}
\end{figure}
 
 Figure \ref{it40} shows the plan of the array at the
 end of January 2009. The inter-station distances
 vary  between 110 and 145 meters.

 \subsection{IceTop triggers and filtering}

 IceTop trigger requires coincidences between six DOMs
 in a time window of 5$\mu$s.
 This is called Simple Multiplicity Trigger (SMT). When this occurs the
 whole detector
 (all IceTop and InIce DOMs) is read. The frequency of IceTop
 SMTs in 2008 was 17 Hz.
 The bandwidth for data transfer from
 South Pole is not large enough to send all triggered events to
 the North. For this reason filtering is required for the event
 fraction that is transmitted. The filtering also checks on the 
 number of stations that participate in the trigger and requires
 at least 3 stations for STA-3 and 8 stations for STA-8 events.
 The frequency of STA-3 events in 2008 was 12 Hz. 
 All coincident triggers between IceTop and InIce SMT (STA-3 + InIceSMT)
 and 5\% of the coincidence events of InIce SMT with surface
 activity (InIce SMT + IceTop) are also transmitted.
 Table~\ref{tab1} shows the different
 filters that were active in 2008.

\begin{table}[htb]
 \vspace*{-10pt}
\caption{Frequency and percentage of different trigger events
 transmitted from South Pole}
 \vspace*{10pt}
\label{tab1}
\begin{tabular}{|l|r|r|}
\hline
 Event type & frequency & \% transmitted \\ \hline
 & & \\
 STA-3 & 12.0 Hz & 20 \\
 STA-8 &  0.5 Hz & 100 \\
 STA-3 + InIce SMT& 2.2 Hz & 100 \\
 InIce SMT + IceTop & 30.7 Hz & 5 \\ 
\hline
\end{tabular}
 \vspace*{-20pt}
\end{table}

\section{IceTop performance}

\begin{figure}[htb]
\vspace*{-10pt}
\begin{center}
\includegraphics[width=7.5truecm]{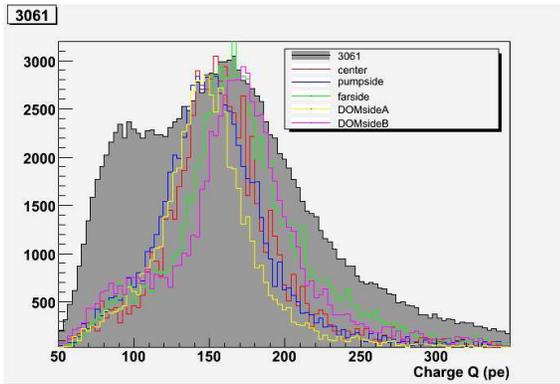}
 \vspace*{-20pt}
\caption{Response of a DOM to muon hits (shaded background)
 and to coincidences with a muon telescope at different locations
 on top of the tank.
\label{fig2}}
\end{center}
 \vspace{-25pt}
\end{figure}

 A very important component of understanding the experimental 
 performance is the calibration of the detector, starting with
 comparison of the signals in high and low gain DOMs and 
 the individual calibration of each DOM.
 DOMs are with different sensitivity that depends on features
 such as the total tank reflectivity, ice quality and 
 temperature (that changes following the outside air temperature).
 For this reason IceTop had special muon calibration
 runs that did not require shower triggers. These runs allow us to
 compare the muon signals in each tank and convert the charge of
 the DOM signal from photoelectrons to VEMs (Vertical Equivalent Muons)
 which makes the whole array response uniform. Fig.~\ref{fig2}
 shows the muon peak in one of the DOMs of station 30. 

 Some tanks are studied with small muon telescopes set in different
 locations on top of the tank. Coincidences between the tank and the 
 muon telescope are used to study the tank response to muons
 hitting the tank at different positions. Five such histograms 
 for DOM 61 of station 30 are also plotted in Fig.~\ref{fig2}.

 One VEM value for this DOM is around 170 photoelectrons
 depending on the exact fit of the distribution.
 The peak of the distribution is slightly higher than those
 of most muon telescope coincidences as the tank is hit by 
 inclined muons that have longer pathlength in the tank and thus 
 generate more light. The use of VEM in the shower analysis and
 the muon calibration makes the detector uniform. 
 A new procedure is currently implemented that will allow us
 to have continuous muon calibration. 

\begin{figure}[htb]
 \vspace*{-20pt}
\begin{center}
\includegraphics[width=8truecm]{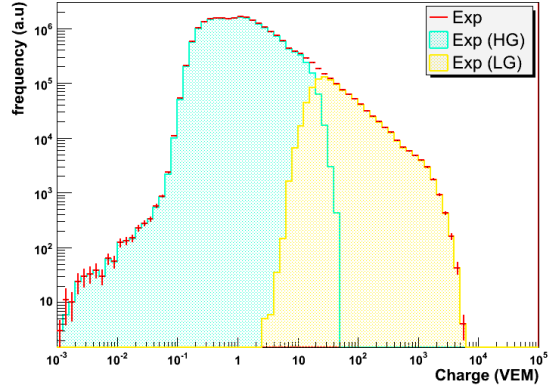}
 \vspace*{-30pt}
\caption{ Frequency of different VEM signal strength reported
 by HG and LG DOMs. Low gain DOMs are responsible for signals
 exceeding 10 VEM.
\label{fig3}}
\end{center}
 \vspace*{-30pt}
\end{figure}

 Another important calibration is that of HG and LG DOMs. It is
 controlled by comparing the signal strength frequency reported
 by HG and LG DOMs which in the ideal case show smoothly
 joining distribution like the one shown in Fig.~\ref{fig3}. 
 
The comparison of the HG and LG DOMs in the same tank is
used for calibration of individual pairs before the overall
distribution shown in Fig.~\ref{fig3} is made.

\section{First results from IceTop}

 The first results on the cosmic ray spectrum were obtained
 by the cosmic ray working group are best
 presented in the PhD thesis of S.~Klepser~\cite{Klepser_PhD,Klepser1}
 from the 2007 IceTop array that consisted of 26 stations.
 This array was fully efficient for proton showers 
 with energy above 1 PeV.

 The work started with a study of the lateral distribution
 of the air shower signal in VEM, which was fitted from
 experimental data as
$$
S(r)\,=\, S_{ref}(r/R_{ref})^{-\beta - \kappa log_{10}(r/R_{ref})}\, ,
$$
 where $r$ is the perpendicular distance from the shower axis,
 $S_{ref}$ is the expected signal at distance $R_{ref}$ and 
 $\beta$  is a slope parameter related to the shower age.
 From simulations $\kappa$ was found to be constant at 0.3.

\begin{figure}[htb]
\begin{center}
\includegraphics[width=78truemm]{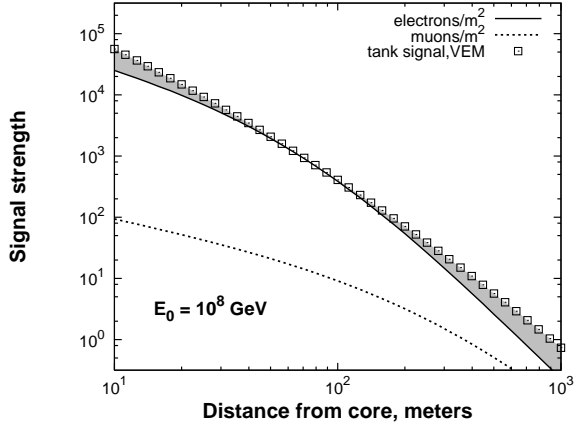}
 \vspace*{-30pt}
\caption{Lateral spread of the shower VEM signal from a proton
 shower in IceTop tanks compared to Greisen's electron and muon
 lateral spread for showers of energy 10$^8$ GeV.
\label{lat}}
\end{center}
\end{figure}

 As shown in Fig.~\ref{lat} this form is quite different from
 the electron density as described in Greisen's formula.
 The reason is that IceTop tanks measure also shower gamma rays
 and muons that that have different lateral spreads from the
 shower electrons.
 
 $S_{ref}$ equal to the average grid of 125 m was used in the
 analysis to relate to the proton primary energy $E_0$.
 A response matrix was calculated to relate the primary and
 reconstructed energy using $S_{125}$ in the case of protons and
 iron showers. At energy of 3 PeV the accuracy in the shower
 core determination was 9 meters and the relative error in
 $log_{10}(E)$ was 0.05 for zenith angles less than 30$^o$.

 Showers in three zenith angle ranges were analyzed. 
 Assuming pure proton or iron composition led to different
 primary spectra in the three angular ranges, while the
 assumption of mixed composition led to a single 
 spectrum with a knee at 3.1$\pm$0.3(stat.)$\pm$0.3(syst.) PeV and
 spectral indices of 2.71$\pm$0.07(stat.) below and 3.11$\pm$0.01(stat.)
 above the knee with systematic uncertainty of 0.08.
 The absolute normalization of the spectrum 
 is lower than in most other measurements.

 This research was conducted when the IceTop detector
 Monte Carlo code was in a very early stage of development.
 The current version of the code is much improved and 
 currently a new analysis is being performed.

 \section{Studies of composition sensitive parameters}

 There are several composition sensitive parameters that can be measured
 by IceCube as a cosmic-ray detector:
 \begin{itemize}
 \item The energy deposited by the shower muon bundle in 1 km of ice.
 Note that at primary energy 10$^{17}$ eV the average muon bundle 
 in iron showers deposits about 2.4 times more energy than that of 
 proton showers.
 \item The waveforms in IceTop tanks can be used for
 a study of the GeV muon density, at least far away from the 
 shower axis.
 \item The lateral distribution of the shower signal is flatter
 for showers of heavy nuclei.
 \item The risetime of the air shower front depends on
 the depth of shower maximum $X_{max}$, 
 and is different for proton and iron showers.
 \item The angular distribution of the showers in IceTop 
 depends on the cosmic-ray chemical composition as the showers
 initiated by heavy nuclei are absorbed much faster as a function
 of zenith angle.
 \item The angular distribution of muon bundles in-ice in
 non-coincident events is also sensitive to the composition
 since proton showers generate higher energy muons that can
 penetrate to the detector at high zenith angles compared
 to those of heavier primaries.
 \end{itemize}
 Our strategy will be to use as many complementary observations
 as possible in order to decrease some of the uncertainties
 inherent in the interaction models needed for simulations and
 interpretation of the data.
 All the parameters listed above will be studied, and the first four, which
 are applicable to IceTop and InIce coincident events, will be 
 determined for each coincident event. One should not forget that
 coincidental events are very close to vertical.    

\vspace{0.3cm}
\noindent
{\bf Angular dependence of showers in IceTop}
As demonstrated by the preliminary analysis~\cite{Klepser1}
the angular dependence of rates on the surface is sensitive to 
composition.  We are in the process of analyzing the full run
of the IceTop 26-station configuration in 2007, which contains
sufficient statistics to extend the analysis above $10^{17}$~eV.
This analysis uses cuts, such as containment of the air shower 
core in the perimeter of the array.

We will continue to use this method for the full IceTop.  It has two
important features, the larger acceptance as compared to coincident
events and its different dependence on models of hadronic interactions.
With the full kilometer squared array we expect sufficient statistics
to cover the 1-2 EeV energy bin.  The method depends on shower absorption
 and is closely related to a measurement of depth of shower maximum.
It is therefore complementary to the analysis with coincident events
that depends on the ratio of the signal of high energy muons to shower size.

\vspace{0.3cm} 
\noindent
{\bf Study of muon bundles in coincident events:}
The acceptance of the telescope formed by IceTop and the
deep IceCube neutrino telescope is approximately 1/3 km$^2$sr,
sufficient to detect tens of events events per year above $10^{18}$~eV.
If Ref.~\cite{HiResMIA} is correct, we may find evidence
suggestive of a transition from galactic to extra-galactic cosmic
rays below this energy.  Although full
acceptance will only be achieved at the end of the deployment
period, already after the 2008/09 season deployment
IceCube is large enough to explore fully the knee region up to
$10^{17}$~eV.  Expected event rates per year (assuming 70\% efficiency)
are indicated in Table~\ref{table}.
\begin{table}[htb]
\vspace*{-10pt}
\caption{Air shower rates in IceCube in 2008 (IT40) and
 upon completion (IT80)}
\vspace*{5pt}
\begin{tabular}{|r|rr|}\hline
E (PeV) & \multicolumn{2}{c}{Rates yr$^{-1}$ln(E)}  \\ \hline
    & IT40 & IT80 \\ \hline\hline
1.0 & $2 \times 10^6$ & $10^7$\\
10. & $3\times 10^4$ & $1.4\times 10^5$\\
100. & 320 & 1600 \\
1000. &  2.5 & 12 \\ \hline
\end{tabular}
\label{table}
\end{table}

\begin{figure}[htb]
\begin{center}
\includegraphics[width=78truemm]{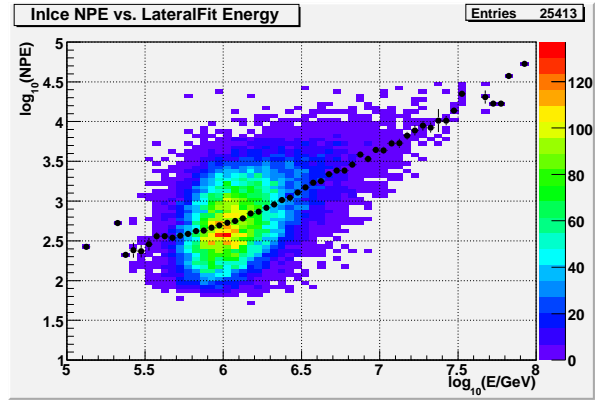}
 \vspace*{-20pt}
\caption{ The total number of photoelectrons (NPE) in IceCube
 is plotted as a function of the reconstructed air shower
 energy in IT40. 
\label{Tilo}}
\end{center}
 \vspace*{-20pt}
\end{figure}

  As an example of what could be achieved by further analysis
 we show in  Fig.~\ref{Tilo} a scatter plot of  the total
 number of photoelectrons (NPE)  InIce versus the shower energy
 reconstructed from the a month of IT40 data. The black dots 
 show the average NPE as a function of energy. The cuts in the
 reconstruction require shower containment on the surface and InIce 
 plus a good reconstruction quality. The NPE is not yet corrected
 for the distance between the DOMs and the bundle trajectory.
 With the 
 statistics shown we can study the composition in the energy
 range of 10$^{5.8}$ to 10$^{6.5}$ GeV. The highest energy events
 collected already approach 10$^8$ GeV.

The data and analysis 
underlying this figure also illustrate the power of coincident
events for energy calibration in IceCube.  Events in which
the shower core is inside IceTop can be reconstructed
and energy assigned with a statistical uncertainty of
order $\delta E / E\,=\,0.1$. To the extent
that the muon content of air showers in the energy range from
$1$ to $100$~PeV are well understood, these coincident events can be
used to calibrate the in-ice signal expected for deposition
of an amount of energy expected from the muon bundle in
the deep array.
 The study of muon bundles InIce is complicated by the very 
 large fluctuations in the high energy muons energy loss.
 
 For downward events, where the signal is dominated by atmospheric muons,
 one can use the muon intensity (to the extent that it is known) to
 estimate the effective attenuation length (a combination of
 scattering and absorption~\cite{Kurt}).  This is done simply by calculating
 the expected intensity of muons at each depth from a standard
 parameterization~\cite{RPP} (accounting for propagation in the
 ice~\cite{LS,MMC}) and fitting the observed DOM counting rate at each depth
 with a parameter that represents the attenuation length of Cherenkov
 light at that depth. We have checked that the effective attenuation
 length derived in this way agrees well with direct measurements~\cite{Kurt}.
 Such a simplified analysis is  useful and fast for simulation of
 downward muon bundles where essentially all events are signal.  

 The detailed study of muon bundles may also help us identify the
 production of very high $p_T$ muons in high energy showers. 

\begin{figure}[htb]
\begin{center}
\includegraphics[width=36truemm]{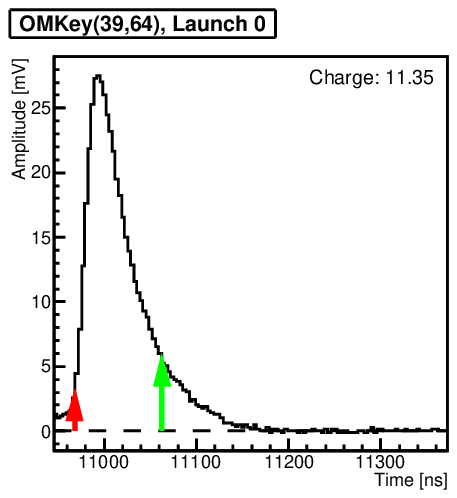}
\includegraphics[width=36truemm]{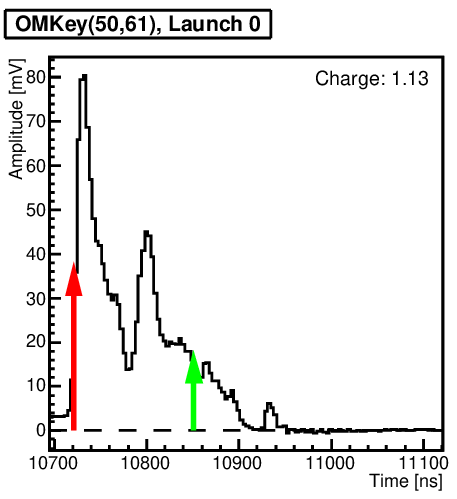}
 \vspace*{-20pt}
\caption{Two waveforms taken at distances of 50 m (left, with 
 a LG DOM) and 250 m from the shower core with a HG DOM.
 The reconstructed shower energy
 is 5$\times$10$^{15}$ eV. Arrows show the leading and trailing
 edges of the waveforms. 
\label{waveforms}}
\end{center}
 \vspace*{-20pt}
\end{figure}

\vspace{0.3cm}
\noindent
{\bf Identifying surface low energy muons from the detected waveforms:}   
 In addition to our principal discriminator of primary composition,
 which is the ratio of shower size at the surface to muon signal
 in the ice, we also have the possibility of recovering some information
 about the relative contribution of GeV muons to the signal
 at the surface, which is also sensitive to composition. 
 Fig.~\ref{waveforms} shows the waveforms of the signals in an air shower
 of reconstructed energy 5$\times$10$^6$ GeV  seen by
 a detector at distance of 50 m from the shower axis and one at
 a distance of 250 m.

 The reconstructed energy of this shower is 5$\times$10$^6$ GeV
 and zenith angle of 40$^o$. The waveform from the tank that is
 closer to the axis contains 11.34 VEM and width consistent with
 the shower front thickness. The waveform at 250 meters from the axis
 contains 1.13 VEM and is much more ragged. It seems possible 
 that a muon hits first the tank and few low energy electrons and photons
 hit the tank later. We will investigate such waveforms coming 
 from Monte Carlo events when the IT40 Monte Carlo produces 
 sufficient statistics. 

 One can in principle deconvolve the waveforms and
 estimate the number of muon peaks contributing to
 the signals in the tanks.  Such a treatment will be most
 effective in large air showers and at a significant distance
 from the shower core where the relative contribution of muons
 to the signal is higher. Thus such an analysis will be 
 most appropriate when a large fraction of IceTop is deployed. 
 Close to the shower core, as in Fig.~\ref{waveforms}, the
 estimate of the muon number may be more difficult.
 The fine timing resolution of the IceCube DOMs (3-4 ns per bin)
 may allow us to identify the GeV muons in air showers
 of primary energy above 10$^{17}$ eV.
  

\end{document}